\newcommand{\eqb}{\begin{equation}}
\newcommand{\eqe}{\end{equation}}
\newcommand{\dmb}{\begin{displaymath}}
\newcommand{\dme}{\end{displaymath}}
\newcommand{\pd}{\partial}
\newcommand{\ep}{\varepsilon}
\newcommand{\eab}{\begin{eqnarray}}
\newcommand{\eae}{\end{eqnarray}}
\newcommand{\ra}{\right\rangle}
\newcommand{\la}{\left\langle}

\documentclass[epj,referee]{svjour}
\usepackage{latexsym}
\usepackage{graphics}
\begin{document}
\title{Magnetic charge of finite lifetime in SU(2) gluodynamics}
\author{Ralf Hofmann}

\institute{Theoretical Physics Institute,  
         University of Minnesota,  
         Minneapolis, MN 55455, USA\\ 
	 present address: Max-Planck-Institut f\"ur Physik, F\"ohringer Ring
	 6, 80805 M\"unchen, Germany}
\date{Received: date / Revised version: date}
%

\abstract{A self-dual, localized solution to the classical SU(2) 
Yang-Mills equation in Euclidean spacetime, which  
formally possesses infinite action, is 
investigated in view of its U(1) charge content 
after Abelian projection. This is suggested by 
noting that the solution satisfies 't Hooft's differential 
projection condition away from the singularities. 
As a result the existence of dynamical, magnetic charge of 
finite lifetime is established. A covariant cutoff for the action is 
introduced by demanding the solution to be close to an 
instanton topologically. This is in analogy to the \underline{calculation} 
of the mass of a point charge in classical electrodynamics or the subtraction of diverging 
self-energies of magnetic monopoles as discussed in the literature. 
The Wilson loop is evaluated in the background of a dilute gas. 
Assuming identical integrals over size distributions, the corresponding 
static quark/anti-quark potential at infinite spatial separation 
can be seizably higher than the potential in a dilute instanton gas.
\PACS{
      {?}{?} \and {?}{?}
     } 
} 

\maketitle

\section{Introduction}

Quantum Chromodynamics (QCD) \cite{QCD} is widely accepted as 
the theory of strong interactions due to its experimentally 
verifyable feature of asymptotic freedom in the UV momentum regime (see for 
example ref.\,\cite{Beneke} and references
therein). A perturbative treatment of the theory in the IR domain is 
excluded by a large coupling constant. Non-perturbative techniques 
are necessary and partial successes in explaining the low energy phenomenology 
of strong interactions, such as spontaneous chiral symmetry 
breaking and the axial U(1) problem, have 
been achieved this way (see ref.\,\cite{Schafer} and references therein). However, 
the basic observation that no free color charges exist in 
nature is understood only poorly so far. 

The dual superconductor picture of 
the non-perturbative QCD vacuum dates 
back to the mid seventies \cite{Mandelstam}. It provides an appealing mechanism for
color confinement due to the formation of color electric 
fluxtubes linking colored objects immersed in a 
condensate of magnetic charges. The resulting potential, linear in the (large) separation 
between the charges, is determined by an effective, universal scale $-$ 
the string tension. The occurence of magnetic charges 
in SU($n$) Yang-Mills theory was made transparent in the process of 
the so-called maximal Abelian gauge fixing (MAGF) \cite{thooft}. Due to this fixing 
gauge transformations belonging to the maximal Abelian 
subgroup U(1)$^{n-1}$ of SU($n$) are unconstrained and the resulting 
Abelian gauge theory has electric and magnetic charges. 
A rather successful model $-$ the dual Ginzburg-Landau theory $-$ 
has been employed to effectively mimic the low energy sector of SU($n$) Yang-Mills 
theory after MAGF \cite{Suzuki}. To classically describe the 
deconfinement phase transition induced by finite temperature 
a non-renormalizable version of the dual Ginzburg-Landau theory  
has been put forward recently \cite{Hofmann}.  

Yet no analytical derivation of magnetic monopole condensation 
has been given although rather impressive results were obtained from 
the lattice \cite{Cherno}. However, the lattice 
detection of monopole trajectories has its inherent uncertainties (see below). Exact 
classical solutions to the Yang-Mills equations subjected to Abelian projection 
have been investigated more recently in view of their 
magnetic charge content \cite{Gubarev,Brower,Reinhardt}. 
Concerning charges of finite lifetime the results are 
not conclusive in the case of instantons since either the gauge fixing 
functional diverges for non-infinitesimal monopole trajectories \cite{Gubarev} or the 
monopole loop is infinitesimal \cite{Brower,Reinhardt}. The 
investigation of field configurations, which are \underline{not} exact 
solutions to the Yang-Mills equations (instanton/anti-instanton pair), are 
usually carried out on the lattice. Since the lattice provides a UV cutoff 
and thereby has a limited resolution and since only approximative 
solutions have been looked at the claimed detection of finite monopole 
loops is not well established.  

The purpose of this work is to point out the 
presence of long-lived magnetic charge due to exact, self-dual solutions 
to the SU(2) Yang-Mills equations and to investigate 
their effect on the static quark 
potential at infinite separation. 
The fact that a solution possessing infinite action has a vanishing contribution 
to the partition function is delt with by an appropriate 
subtraction in the spirit of ref.\,\cite{thooft2}. The divergence of the action is traced back to singularies of the action density on 
a hypersurface. In ref.\,\cite{thooft} the appearance of magnetic 
monopoles in SU($n$) gauge theory was made manifest after imposing a 
non-propagating gauge which leaves the maximal Abelian subgroup 
U(1)$^{n-1}$ invariant. This gauge condition demands that a matrix 
quantity, say the field strength component $F_{12}$, 
transforming homogeneously 
under gauge transformations, be diagonal. In particular, for the case of SU(2) it 
is easy to shown that at points, where this condition is 
no constraint on the corresponding gauge transformations, the inhomogeneous part of 
the gauge transformation of $A_\mu$ is singular 
and contains an Abelian, magnetic monopole \cite{Toki}. 
Due to the Coulomb-like behavior of the magnetic field strength 
in the vicinity of such a point the action of the field configuration 
formally diverges, and hence it must be regularized phenomenologically. 
The solution discussed obeys the 
differential Abelian projection condition away from the singularities, and hence the restriction to its  
diagonal component is suggested.      

In the present work magnetic charge is identified by means of the 
Coulomb-like behavior of the field strength of a fictitious charge moving toward (away from) 
a spatial point which turns into a singularity at a certain pair of time slices. 
The definition of the physical action is performed 
in analogy to defining the mass of a static point charge 
in classical electrodynamics. A covariant cutoff for the action 
functional is introduced which, in principle, 
can be determined phenomenologically. 

Summarizing, the line of reasoning 
in this work is as follows: Classical, self-dual, infinite action solution, which is regular 
almost everywhere, possesses a
converging action integral 
in the infinite, and is in Abelian gauge (Section 2) $\to$ presence of magnetic charge $\Rightarrow$ 
infinite action (Section 3) $\to$ phenomenological action regularization (Section 4). 
In addition, the static quark/anti-quark potential in a 
dilute gas is calculated for infinite spatial separation in Section 5. Thereby, 
the integral over size distribution is assumed to be 
identical to that of the instanton case. The result is compared 
to the potential due to a dilute instanton gas.

\section{Classical solution}

A class of winding, BPS staturated solutions to the SU(2) 
classical Yang-Mills equations in Euclidean spacetime was found a 
long time ago in ref.\,\cite{BPST}. These solutions have finite classical 
action $S=1/(4g^2)\int d^4x 
F^a_{\mu\nu}F^a_{\mu\nu}=8\pi^2/g^2$ and 
spontaneously break eight of the symmetries of the classical theory. They 
are (anti)self-dual, that is
\eqb
\label{sd}
F^a_{\mu\nu}=\pm\ep_{\mu\nu\kappa\lambda} F^a_{\kappa\lambda}\ , 
\eqe
and are known as instantons. As first suggested by Gribov 
instantons can be interpreted as tunneling 
trajectories linking topologically distinct vacua \cite{Gribov}. In the last 20 years or so 
there has been an enormous industry relating the presence of 
these solutions in the non-perturbative QCD vacuum to 
low energy phenomenology, namely 
the $U(1)_A$ problem and spontaneous chiral symmetry breaking (see ref.\,
\cite{Schafer} and references therein). However, 
the issue of color-confinement, which should be an important 
consequence of QCD, has not been satisfactorily 
settled by the instanton calculus 
\cite{Gross,Diakonov2}. 

For gauge group SU(2), in singular gauge, 
and for center $z=0$ the instanton solution is given as 
\eqb
\label{is}
A^a_\mu=2\bar{\eta}_{a\mu\nu}\rho^2\frac{x_\nu}{x^2(x^2+\rho^2)}\ , 
\eqe
with $\bar{\eta}_{a\mu\nu}\equiv-\frac{i}{2}\,\mbox{tr}(\tau_a \tau_\mu^{-}\tau_\nu^{+})$, 
$\tau_\mu^{\pm}=(\vec{\tau},\mp i)$, and $\tau_a$ the Pauli matrices. 
The solution (\ref{is}) can 
be obtained as follows \cite{Diakonov}. Making the ansatz  
\eqb
\label{ans} 
A^a_\mu=\bar{\eta}_{a\mu\nu}x_\nu\frac{1+\phi(x^2)}{x^2}\ ,
\eqe
the action takes the form
\eqb
\label{actphi}
S=\frac{8\pi^2}{g^2}\frac{3}{2}\int d\tau\left\{
\frac{1}{2}(\phi^\prime)^2+\frac{1}{8}(\phi^2-1)^2\right\}\ ,\ \ 
\tau\equiv\ln\left(\frac{x^2}{\rho^2}\right)\ ,
\eqe
where $\rho$ is some length scale. A BPS saturated vacua interpolation of 
the theory (\ref{actphi}) is
\eqb
\label{sol}
\phi=-\tanh\left(\frac{\tau}{2}\right)\ .
\eqe
Substituting this into eq.\,(\ref{ans}), one 
obtains the solution of eq.\,(\ref{is}). Indeed, evaluating the action yields 
\eqb
S=\frac{8\pi^2}{g^2} 
\frac{3}{8}\int d\tau \cosh^{-4}(\frac{\tau}{2})=\frac{8\pi^2}{g^2}\ .
\eqe
Another BPS saturated solution to the equation of motion corresponding to 
(\ref{actphi}) is obtained by letting $\tau\rightarrow\tau+\tau_0$ in (\ref{sol}). 
Thereby, $\tau_0$ may be complex as long as the new solution is real. This is necessary 
to have a real action. Admissible complex shifts are $\tau_0=in\pi$ with $n\in{\bf Z}$. 
Choosing $\tau_0=i\pi$ yields
\eqb
\label{sol2}
\phi\rightarrow-\coth\left(\frac{\tau}{2}\right)\ .
\eqe
Note that (\ref{sol}) and (\ref{sol2}) behave identically in the infinite. 
For the Yang-Mills potential eq.\,(\ref{sol2}) implies
\eqb
\label{sph}
A_\mu\to A_\mu=-2\bar{\eta}_{a\mu\nu}\rho^2\frac{x_\nu}{x^2(x^2-\rho^2)}\ .
\eqe
The corresponding anti-solution (negative topological 
charge density) is obtained by substituting $\bar{\eta}_{a\mu\nu}\to \eta_{a\mu\nu}$. 
Formally, the action of solution (\ref{sph}) reads
\eqb
S=\frac{8\pi^2}{g^2} 
\frac{3}{8}\int d\tau \sinh^{-4}(\frac{\tau}{2})\ .
\eqe
It diverges due to the quartic pole of the integrand at $\tau=0$ and \underline{not} 
due to a diverging integral in the infinite. An infinite action of a 
classical solution generally is sufficient cause to disregard 
this solution on grounds of its vanishing
contribution to the partition function of the theory. As it will be explained below there 
is physical information in the solution (\ref{sph}), and 
hence it will not be dismissed.  

The surface of divergence of (\ref{sph}) is the 3-sphere $x^2=\rho^2$. For this reason it is refered to 
as {\em spheron} throughout the remainder of the paper. Writing the spheron as
\eqb
\label{sper}
A^a_\mu=-\bar{\eta}^a_{\mu\nu}\pd_\nu \log\left(1-\frac{\rho^2}{x^2}\right)\ ,
\eqe
it is obvious that away from the singularities 
the following two sets of conditions are fulfilled
\eqb
\label{gauges}
\left(\pd_\mu\pm i A^3_\mu\right)A^\pm_\mu=0\ ,\ \ \ \ \ \pd_\mu A^a_\mu=0\ ,
\eqe
where $A^\pm_\mu\equiv A^1_\mu\pm i A^2_\mu$. 
The first set expresses the fact that the spheron is in 
't Hoofts maximal Abelian gauge (MAG)\cite{thooft} which leaves 
a U(1) subgroup of SU(2) unfixed. Globally, this is mirrored by 
a minimum of the functional 
\eab
\label{GM}
G_M[A]&=&\frac{1}{4}\int d^4x \left\{A^1_\mu(x)A^1_\mu(x)+A^2_\mu(x)A^2_\mu(x)\right\}\nonumber\\ 
&=&2\int d^4x\, \frac{\rho^4}{x^2(x^2-\rho^2)^2}
\eae
under gauge transformations \cite{Brower} if 
the divergence on $x^2=\rho^2$ is cut out (see Section 4). 
Obeying of the second set of conditions in (\ref{gauges}) indicates 
that the spheron is Lorentz gauged. 
Restricting oneself to the diagonal component, which is suggested by MAG and Abelian 
dominance (see for example \cite{Suzuki2}), the Lorentz gauge guarantees the minimization of the functional 
\eqb
\label{GL}
G_L[A]=\frac{1}{4}\int d^4x\, A^3_\mu(x)A^3_\mu(x)=\int d^4x\, \frac{\rho^4}{x^2(x^2-\rho^2)^2}
\eqe
under U(1) gauge transformations \cite{Zakharov}. 

Back to the problem of a diverging action: It is argued in the 
following that our poor understanding of the nature of charged particles 
in classical electrodynamics must necessary carry over to 
a classical, Abelian projected SU(2) (and more general SU($n$)) gauge 
theory in which the 
appearance of localized U(1) charges is 
expected \cite{Mandelstam,thooft}. The term "poor understanding" 
refers to the fact that the classical
self-energy of a static point charge is 
divergent. In classical electrodynamics the usual procedure 
is to constrain the corresponding energy functional to regions of space  
with a regular behavior of the integrand to 
produce the \underline{observed} mass of the associated particle. In the 
quantum theory of electrodynamics (QED) the mere introduction 
of a mass parameter 
into the Lagrangian accounts in an effective fashion 
for the aforementioned process in the limit of zero coupling 
\footnote{In principle, the mass of the particle can then still be determined gravitationally.}. 
At nonzero coupling the mass starts to run (albeit slowly in QED due to 
an IR conformal point) with the momentum probing it. Going from the IR regime up to large momenta 
(yet below the Landau pole) the gross contribution to the mass is 
of a classical origin as the weak logarithmic dependence 
on momentum shows. From the viewpoint of 
Abelian projection it is thus 
\underline{not} imperative to disregard a classical solution of 
infinite action due to divergences of the action density on 
hypersurfaces if these divergences 
can be attributed to U(1) charge in the 
Abelian projected theory (see ref.\,\cite{thooft2} 
where the subtraction of an infinite self energy 
was made explicit in the case of 2+1 dimensional QCD).

\section{Magnetic charge}

In this Section the consequences of Abelian 
projection applied to the spheron are investigated. For fixed 
times $|x_4|\le\rho$ there is a 2-sphere of 
divergence with radius $\tilde\rho(x_4)=\sqrt{\rho^2-x_4^2}$. At $x_4=-\rho$ a pointlike 
singularity in 3-space emerges which starts expanding with Minkowski-space 
velocity $v(x_4=-\rho)=1/\sqrt{2}$. The 2-sphere reaches 
its maximal radius $\tilde\rho=\rho$ at $x_4=0$ with velocity $v(x_4=0)=0$, and 
shrinks to a point at $x_4=+\rho$. At times $|x_4|>\rho$ there is no 2-sphere, 
and the field is regular in 3-space. 

The underlying theory is matter-free 
and non-Abelian. The luxury of placing a \underline{static} $\delta$-source on 
the right-hand side of Maxwell's equation to obtain a Coulomb law for 
the corresponding field strength is absent. If attributed to 
the spatial divergences of the spheron field 
the life of a U(1) charge is highly dynamic and of limited duration. Hence, there are strong 
retardation and threshold effects which 
necessarily blurr the Coulomb picture. Moreover, 
in the case of accelerated magnetic charge the conservation 
of magnetic flux can be realized by a 
smeared flux distribution 
in contrast to the static Dirac string. Thus, at $x_4=\pm\rho$ there may be 
divergences of $\vec B^2$ 
higher than quartic in $|\vec x|$. It is then not justified to read 
off the magnetic charge 
from the "residue" belonging to a
fourth-order Coulomb pole $|\vec x|=0$ of 
$\vec B^2$ at $x_4=\pm\rho$. However, if there is charge a Euclidean 
space observer placed at $|\vec x|=0$ measures squares of field strengths 
that are quartically divergent as $x_4\to-\rho$. Close to the singularity 
this behavior can be attributed to a fictitious charge moving \underline{uniformely} at 
Euclidean "speed-of-light" along one of the spatial coordinate axis 
toward the observer\footnote{The Lorentz boost 
connecting the rest-frame of this charge to that of the observer is a 
O(4) rotation $T$ of angle $\frac{\pi}{2}$. 
For a charge moving along the 1-axis one has
\eqb
T=\left(\begin{array}{cccc} 0 & 1 & 0 & 0\\ 
-1 & 0 & 0 & 0\\ 
0 & 0 & 1 & 0 \\ 
0 & 0 & 0 & 1\end{array}\right)\ .
\eqe
It is easy to check that the boost $T$ leaves the 
field strength components $E_1$ and $B_1$ invariant.}. 
At $x_4=-\rho$ the hypothetical charge materializes, that is, there is a 
genuine singularity at $\vec x=0$. After Abelian projection the squares of 
electric and magnetic charges can thus be read off as the "residues" of the 
fourth-order Coulomb pole at 
$x_4=-\rho$ ($\vec x=0$) in $\vec E^2$ and $\vec B^2$, respectively. 
At $x_4=\rho$ ($\vec x=0$) a similar argument applies. 
The situation is summarized in Fig. 2.

Using eq.\,(\ref{sper}) and the 
properties of the $\eta$-symbols, one easily obtains
\eab
\label{E^2}
\vec E^2&\equiv&\left(\pd_4 A^3_i-\pd_i A^3_4\right)^2=\left(\pd_4A^3_i\right)^2-
2\pd_4A_i\pd_iA^3_4+\left(\pd_iA^3_4\right)^2\nonumber\\ 
&=&\left(\pd_1\pd_4h(x^2)\right)^2+\left(\pd_2\pd_4h(x^2)\right)^2+\left(\pd_4^2h(x^2)\right)^2+\nonumber\\ 
& &\left(\pd_1\pd_3h(x^2)\right)^2+\left(\pd_2\pd_3h(x^2)\right)^2+\left(\pd_3^2h(x^2)\right)^2-\nonumber\\ 
& &-2\pd_4^2h(x^2)\pd_3\pd_4h(x^2)\ ,\ \ \ \ \ \ \ h(x^2)\equiv\log(1-\frac{\rho^2}{x^2})\ .
\eae
For $x_4\to\pm \rho$ and $\vec x=0$ eq.\,(\ref{E^2}) 
reduces asymptotically to 
\eqb
\label{ech}
\vec E^2_{x_4\to\pm\rho,\vec x=0}=\frac{1}{|x_4\mp\rho|^4}\ .
\eqe
Converting from the non-perturbative to 
the perturbative definition of $A_\mu$ by virtue 
of $A_\mu\to \frac{1}{g}A_\mu$ and comparing with the standard Coulomb expression, 
the magnitude of the electric charge $e$ is read off as 
\eqb
\label{e}
|e|=\frac{4\pi}{g}\ .
\eqe
For $\vec B^2$ one finds
\eqb
\vec E^2\equiv\left(\ep_{ijk}\left[\pd_j A^3_k-\pd_k A^3_j\right]\right)^2=
4\left((\pd_mA^3_n)^2-\pd_mA^3_n\pd_nA^3_m\right)\ ,
\eqe
where the expression in terms of derivatives of 
$h(x^2)$ has been omitted. The "residue" of the Coulomb pole $m$ turns out to vanish. 
Eqs.\,(\ref{e}) signals that $e$ fulfills  
the Dirac condition with respect to the SU(2) coupling $g$. Thus it should be interpreted as a 
magnetic charge\footnote{The corresponding quark doublet would have electric 
charge $\pm\frac{1}{2}g$ \cite{thooft}.}. This can be made manifest by exploiting the 
following charge rotation symmetry of Maxwell's equations \cite{thooft}
\eab
m&\to& m\cos\alpha+e\sin\alpha\ ,\nonumber\\ 
e&\to& -m\sin\alpha+e\cos\alpha\ , 
\eae
which allows to rotate $e$ into $m$ for $\alpha=\frac{\pi}{2}$. 
In Fig. 3 the angular distribution of $\vec E^2$ at three different instants is depicted. 

The concentration of field strength in the "northern" hemisphere at $x_4=-\rho$ evolves
to a mirror symmetric distribution at $x_4=0$. At $x_4=\rho$ field 
strength is concentrated in the "southern" 
hemisphere. 

At this point it is worth comparing the above results to the work of 
ref.\,\cite{Brower}. There, an infinitesimal magnetic 
monopole loop was obtained by looking at singular gauge 
transformations applied to an instanton in singular gauge. These transformations 
were parametrized 
by a radius scale $R$ deviding 
spacetime into "outer" and "inner" regions, with 
the gauge transformed 
instanton in singular and 
regular gauge, respectively. The MAG functional 
turned out to be stabilized at $R=0$, that is, back at the instanton. 
Now, the instanton has finite action. Does this contradict the things said 
before? It does not because the identified 
magnetic charge lives only infinitesimally short in such a 
way as to keep the action finite. In ref.\,\cite{Brower} 
further investigations toward the detection of finite-size 
monopole loops were carried out on the lattice 
for an instanton/anti-instanton (I-A) pair. Since the lattice naturally provides a UV cutoff field strength 
divergences remain hidden. Furthermore, the I-A system is not an exact solution to the
equation of motion and the results of an abelian 
projection may heavily depend 
on the chosen ansatz for a 
variational principle approach\footnote{In the streamline approach there is no 
stabilization at finite distance between the 
centers \cite{Yung} whereas in the simple sum ansatz there is \cite{Diakonov1}.}. 
Thus, the claimed detection of monopole loops in the I-A system using 
the lattice \cite{Brower} remains questionable.

\section{Regularized action}

In this Section a covariant definiton of the physical spheron action is suggested. 
For the spheron to be of relevance its contribution to the partition function 
(or tunneling amplitude in Minkowski space) 
must be seizable. Moreover, its topological charge is up to the factor $8\pi^2/g^2$ equal 
to the action (for the anti-spheron equal minus the action) 
and is strictly 
positive (negative) for finite cutoff. In classical electrodynamics the cutoff for 
the energy functional of the electron's Coulomb field is 
determined from its measured mass. Similarily, estimating the mean topological susceptibility 
of the vacuum $\la\ep_{\mu\nu\kappa\lambda} F^a_{\kappa\lambda} F^a_{\mu\nu}\ra$ in terms of the 
measured masses $m_{\eta^\prime}, m_K, m_\pi, m_\eta$ by means of 
the Veneziano-Witten relation \cite{VenWit} or estimating the gluon condensate by 
means of QCD sum rules \cite{SVZ}, one could, in principle, approximate the action 
cutoff for the spheron phenomenologically. For now the covariant cutoff $\tau_m$ 
is simply assumed to produce 
the classical action $S=8\pi^2/g^2$ of the instanton. The cut off 
action integral for the spheron is evaluated as
\eab
S_s&=&\frac{8\pi^2}{g^2}\left(\frac{1}{2}-\frac{1}{8}\left[
\cosh(3\tau_m/2)-3\cosh(\tau_m/2)\right]\sinh^{-3}(\tau_m/2)\right)\ ,\nonumber\\   
\tau_m&\equiv&\log(x^2_m/\rho^2)\ .
\eae
Setting $S_s=8\pi^2/g^2$, one obtains $\tau_m=\log3$. In units of $\rho$ the minimal distance $\xi$
to the sphere consequently is $\xi\equiv\sqrt{\exp(\tau_m)}-1\sim 0.73$. For this particular value 
the MAG functional of eq.\,(\ref{GM}) yields 
\eqb
G_M[A]=4\pi^2\rho^2\frac{1}{\exp(\tau_m)-1}=2\pi^2\rho^2
\eqe
as opposed to $G_M[A]=4\pi^2\rho^2$ for the instanton in singular gauge. In Section 5 the 
cutoff will be varied to test the sensitivity of the static quark potential.

\section{Wilson loop in a dilute spheron gas}

In this Section the static quark/anti-quark potential $E(R)$ is 
estimated from the rectangular Wilson loop\footnote{Time-like lines of integration separated by spatial 
distance $R$ and running from $-\infty$ to $+\infty$.} in a dilute spheron gas background 
and compared to that of a dilute instanton gas 
for infinite spatial separation $R\to\infty$. All 
conventions and general results, which do only exploit the 
localization property of the instanton in singular gauge, can be taken from ref.\,\cite{Gross}. 
There, $E(R)$ was obtained as
\eqb
\label{En}
E(R)=-\int d^3x \frac{d\rho}{\rho^5} D_n(\rho)
\frac{1}{n}\mbox{tr}\left(U^+(\vec x) U^-(\vec x+\vec R)-1\right)\ ,
\eqe
where 
\eqb
\label{U}
U^\pm\equiv{\cal P}\mbox{exp}\left(\pm i\int dx_4  A_4(\vec x)\right)\ ,
\eqe
$\cal P$ demands path-ordering, $D_n(\rho)$ denotes the spheron size distribution in a dilute gas for 
gauge group SU($n$), where 
quantum fluctuations are included in a semiclassical way. 
For $R\to\infty$ one has $U^-(\vec x+\vec R)\to +1$. In this limit 
eq.\,(\ref{En}) reduces to\footnote{Note that the authors of 
ref.\,\cite{Gross} work in regular gauge whereas here the spheron behaves like an 
instanton in \underline{singular} gauge in the infinite, thus $\lim_{R\to\infty}
U^-(\vec x+\vec R)=+1$ and \underline{not} $-1$ as in \cite{Gross}.}
\eab
\label{Ene}
\lim_{R\to\infty}E_s(R)&=&-\int d^3x \frac{d\rho}{\rho^5}\, D_n(\rho)\,
\frac{1}{n}\,\mbox{tr}\left(U^+(\vec x)-1\right)\nonumber\\ 
&\equiv&
-\int d^3x \frac{d\rho}{\rho^5}\, D_n(\rho)\,
\frac{1}{n}\,f_s(u)\ ,
\eae
where $u\equiv\sqrt{\vec x^2/\rho^2}$. It is \underline{assumed} here 
that the integral over spheron and instanton size distributions 
are identical\footnote{The spheron 
cutoff is close to $\rho$ for 
topological charge of order one. A constant factor not too far from 
unity distinguishes instanton and spheron distributions since there are slightly 
different scales involved and also due to the low lying non-zero mode 
eigenvalues of gaussian operators in the respective 
backgrounds being different (one-loop contribution to tunneling amplitude \cite{thooft1,Gross}). 
There is no easy estimate of these effects, 
and one would have to perform the one-loop calculation directly 
along the lines of ref.\,\cite{thooft1}. Furthermore, the cutoff for the size 
integrations may phenomenologically 
turn out to be different for the instanton as opposed to the spheron case.}. 
Implementation of the covariant cutoff $\tau_m$ and computation of the 
$x_4$ integral of eq.\,(\ref{U}) yields
\eab
& &f_s(u)=\nonumber\\ 
& &-\mbox{tr}
\,\left(\mbox{exp}\left(-2i\rho^2 \vec\sigma\cdot\vec x\int_{\sqrt{\rho^2\ep_m-\vec x^2}}^\infty
dx_4\,\frac{1}{(x_4^2+\vec x^2)(x_4^2+\vec x^2-\rho^2)}\right)-1\right)\nonumber\\ 
&=&-2\left(\cos\left\{\pi-2\arctan\left(\frac{\sqrt{\ep_m-u^2}}{u}\right)\right\}+
\right.\nonumber\\ 
& &\left.\theta(u-1)\times\cos\left\{\frac{u}{\sqrt{u^2-1}}\left[\pi+2\arctan
\left(\sqrt{\frac{\ep_m-u^2}{u^2-1}}\right)\right]\right\}+\right.\nonumber\\  
& &\left.\theta(1-u)\times\cos\left\{\frac{u}{\sqrt{1-u^2}}\log\left(\frac{\sqrt{\ep_m-u^2}-\sqrt{1-u^2}}
{\sqrt{\ep_m-u^2}+\sqrt{1-u^2}}\right)\right\}-1\right)\ .
\eae
Using $\ep_m\equiv\exp(\tau_m)=3$ from Section 4 and performing the remaining space 
integral of eq.\,(\ref{Ene}), one has
\eab
\label{Ec}
\lim_{R\to\infty}E_s(R)&=&-\int \frac{d\rho}{\rho^2}\, \frac{1}{n}D_n(\rho)\,
\int_{0}^{\sqrt{\ep_m}}du\, u^2 f_s(u)\nonumber\\ 
&\sim&\int \frac{d\rho}{\rho^2}\, \frac{1}{n}D_n(\rho)\times 8\pi\times 1.52\ .
\eae
The result for the instanton in singular gauge is
\eqb
\label{fi}
f_i(u)\equiv-\mbox{tr}\left(U^+-1\right)=2\left(\cos\left\{\pi\left[1-
\frac{u}{\sqrt{1+u^2}}\right]\right\}-1\right)\ ,
\eqe
which implies
\eab 
\label{Ei}
\lim_{R\to\infty}E_i(R)&=&-4\pi\int \frac{d\rho}{\rho^2}\, \frac{1}{n}D_n(\rho)\,\nonumber\\ 
\int_{0}^{\infty}du\, u^2 f_s(u)&\sim&\int \frac{d\rho}{\rho^2}\,\frac{1}{n} D_n(\rho)\times 8\pi\times 1.10\ .
\eae
Hence, for a cutoff chosen to give spheron and instanton identical topological 
charge the $\rho$-independent parts of the integrands are comparable. Setting $\xi=1$ and thereby avoiding 
the problem of slightly different scales 
in the cut off solution, the number 1.52 in eq.\,(\ref{Ec}) changes to 2.41. For even 
higher cutoff, $\xi=\sqrt{5}-1\sim 1.24$, one obtains 3.52. 
Depending on the phenomenologically favorable cutoff this might have 
implications for color confinement would it qualitativly carry over 
to the spheron liquid. In ref.\,\cite{Diakonov2} 
the static quark/anti-quark potential in the 
instanton liquid was evaluated for two average 
instanton sizes $\bar{\rho}$. The values $\bar{\rho}=600\,$MeV$^{-1}$ and 
$\bar{\rho}=400\,$MeV$^{-1}$ yielded $\lim_{R\to\infty}E_i\sim 140\,$MeV 
and $\lim_{R\to\infty}E_i\sim 460\,$MeV, respectively. This asymptotic behavior is approached 
rapidly (at $R\sim 2\,$fm one has $E_i\sim 110\,,$MeV $\bar{\rho}=400\,$MeV$^{-1}$). 
Concentrating on the case $\xi=1$, 3.2 times these 
values (instanton plus spheron contributions) are $\lim_{R\to\infty}(E_s+E_i)\sim 450\,$MeV and 
$\lim_{R\to\infty}(E_s+E_i)\sim 1.5\,$GeV which includes the range of 
spectral continuum thresholds $\sqrt{s_0}\sim 1.2\dots1.4\,$ GeV 
successfully used in light-quark channel QCD sum rules \cite{SVZ}. 
This means that the out-of-the-vacuum creation of a light pair of 
current quark and anti-quark eventually turning 
into the massive constituents of hadrons (conceivably not without exciting more 
current quark/anti-quark pairs) would happen to a good approximation in the 
asymptotically free regime and 
thereby be supported by a large phase space. In this picture the notion of absolut 
quark confinement due to an infinite static 
quark potential for infinite separation 
seems too restrictive. It would be sufficient 
to demonstrate that the integrated probability of the aforementioned pair 
creation process is practically unity  
for separations comparable to or smaller than 
typical hadron sizes. 
At this point the somewhat speculative nature of the above 
considerations is stressed: Apart from the 
question of how physical the assumption is that the spheron's topological charge 
takes values around unity it is necessary to calculate 
the numerical factor distinguishing 
instanton and spheron size distributions for decisive statements. Moreover, 
it is not quite clear how much the interaction 
between anti-instantons with spherons (and vice versa) and 
the interaction between spherons and anti-spherons 
would alter what was said above. It is hoped that these uncertainties 
can be positively eliminated in the near future. At this stage the above results 
still should be taken as another  
serious indication that the physics of color confinement 
may be linked to the 
existence of long-lived magnetic charges 
in the vacuum of the corresponding SU($n$) gauge theory.

\section{Summary}

In this work the identification of long-lived magnetic charges in the 
the vacuum of SU($2$) Yang-Mills theory has been persued. 
The existence of a self-dual, infinite 
action solution to the classical Yang-Mills 
equation, which obeys the differential Abelian 
projection conditions away from its singularities, proved essential 
for this identification. Furthermore, the Abelian component of this 
solution is in Lorentz gauge which suggests an 
interesting connection to the recently discussed 
appearance of a mass dimension two 
condensate in QCD \cite{Zakharov}. For the solution to 
contribute to the partition function in an essential way 
its physical action must be small. In analogy to the definition of the 
mass of a static point charge in 
classical electrodynamics a covariant 
cutoff for the action has been introduced. In this work it was chosen 
such as to give topological 
charge of order one to the solution. The Wilson loop 
in a dilute gas was evaluated, and 
the static quark/anti-quark potential for infinite spatial 
separation was extracted. It turned out that this number 
can be seizably higher than 
the one obtained in a dilute instanton gas 
if identical integrals over size distributions 
are assumed. 

The question whether one can describe low energy 
phenomena in QCD in a semi-classical way 
starting from its classical Lagrangian remains. 
It is well possible that the influence of quantum fluctuations generates a quantum 
effective action at low scales which is even 
qualitatively different from the classical theory. However, the
hope is there that at least the relevant degrees of freedom in the low energy domain 
can be identified in the classical theory.

\section*{Acknowledgements}

The author thanks G. Gabadadze and M. Pospelov for 
stimulating discussions at an early stage of this
work. Useful conversations with T. ter Veldhuis are appreciated. 
The fertile atmosphere at Theoretical Physics Institute and 
the hospitality extended to the author, in particular by M. Shifman,  
are gratefully acknowledged. The author thanks Karin Thier for her 
understanding and her endurance during the past year's 
hardships of spatial separation. This work was funded by Deutscher 
Akademischer Austauschdienst (DAAD).

\noindent
\newpage \vspace*{1cm}
\noindent Figure 1:\ {The integration contours for the formal action functional.}\vspace{1cm}\\ 
\noindent Figure 2:\ {The Euclidean world-volume of singular magnetic flux 
due to the spheron field after Abelian projection.}\vspace{1cm}\\ 
\noindent Figure 3:\ {Angular distribution of $\vec E^2$ at $x_4=-\rho$ (a), $x_4=0$ (b), 
and $x_4=\rho$ (c) for $\rho=1$ and spatial distance 
from the singularity $\zeta=0.5$.} 

\newpage
\newpage
\begin{figure}
\vspace{6.2cm}
\includegraphics{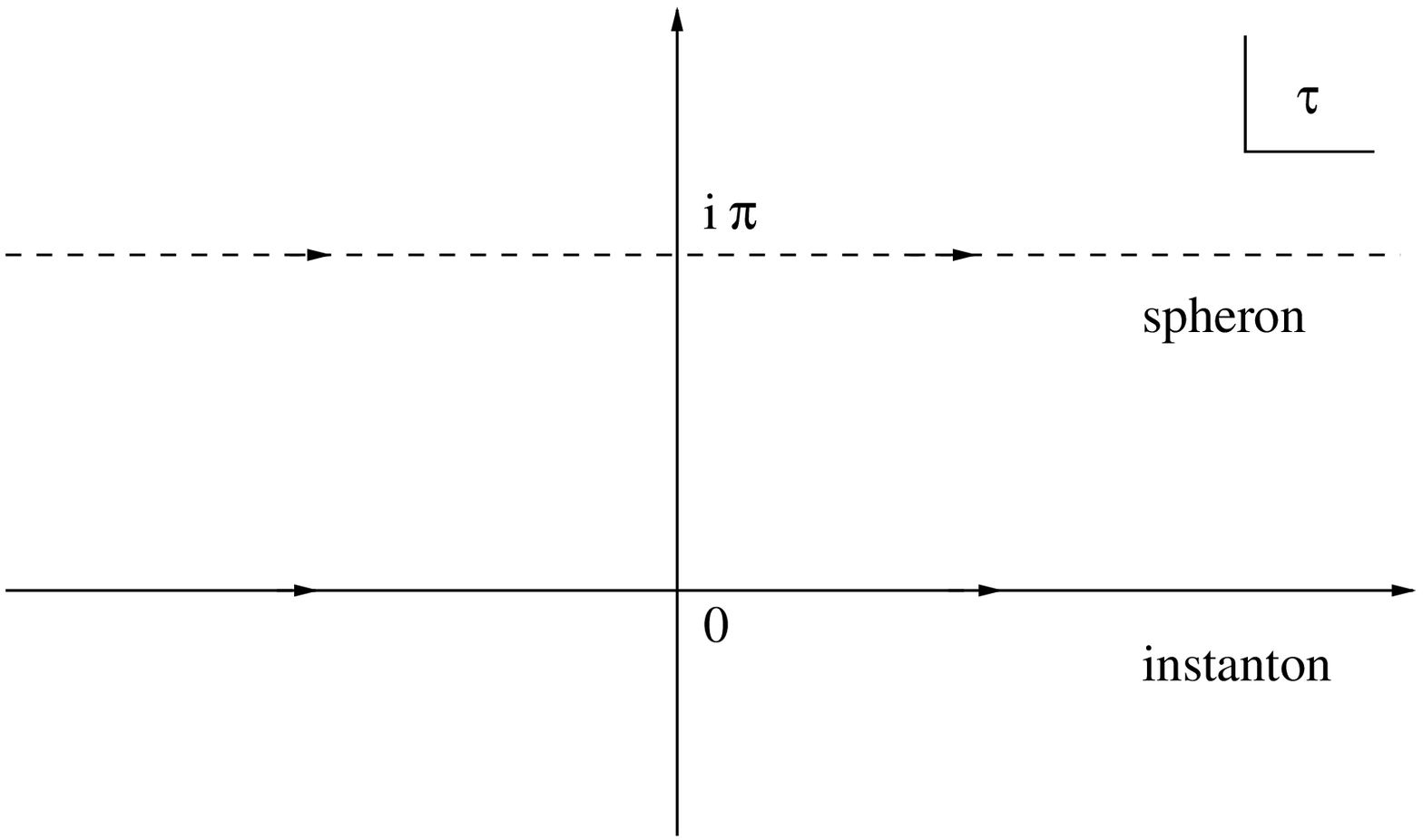}
\caption{} 
\label{} 
\end{figure}
\begin{figure}
\vspace{6.2cm}
\includegraphics{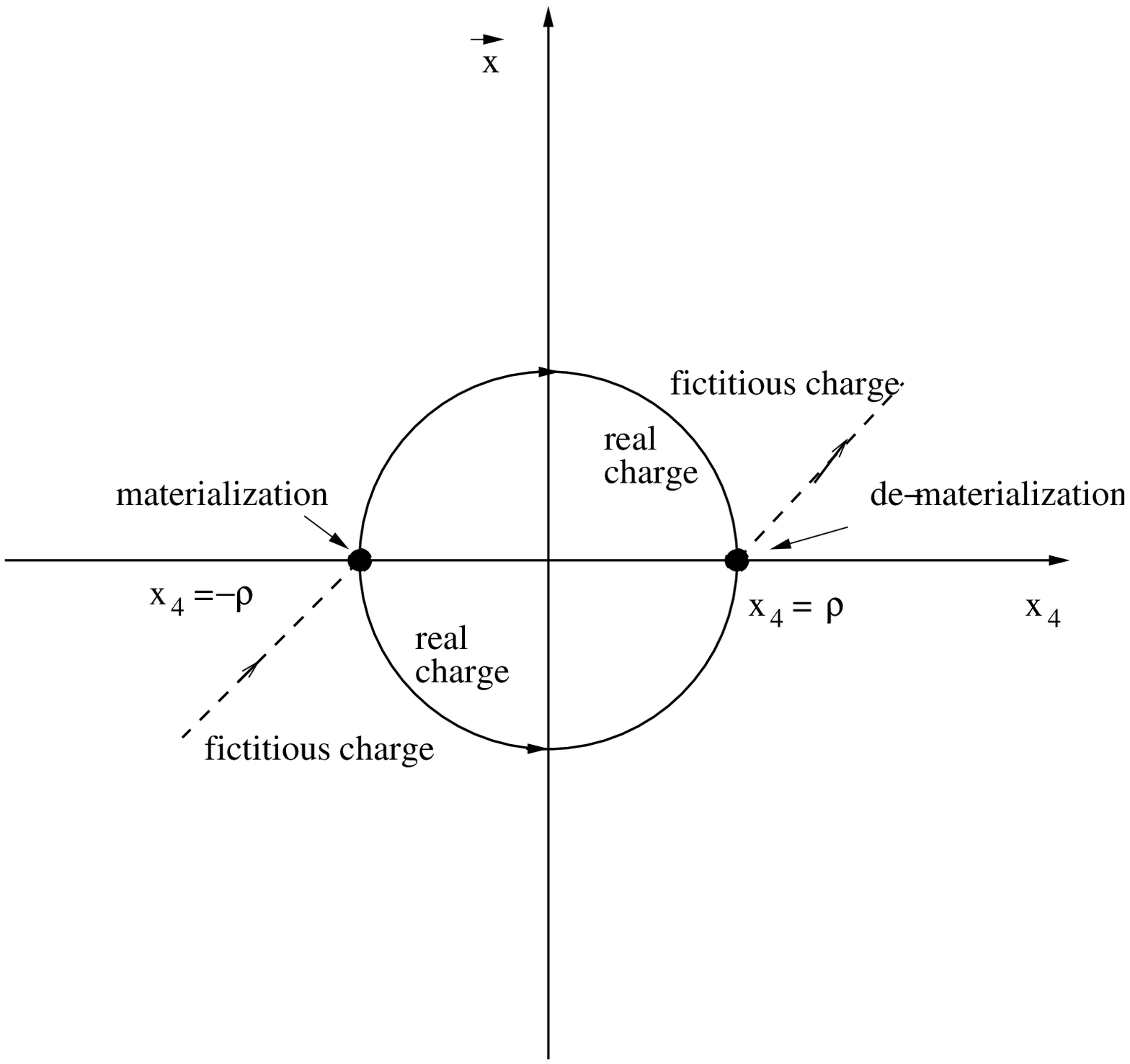}
\caption{} 
\label{} 
\end{figure}
\begin{figure}
\vspace{6.2cm}
\includegraphics{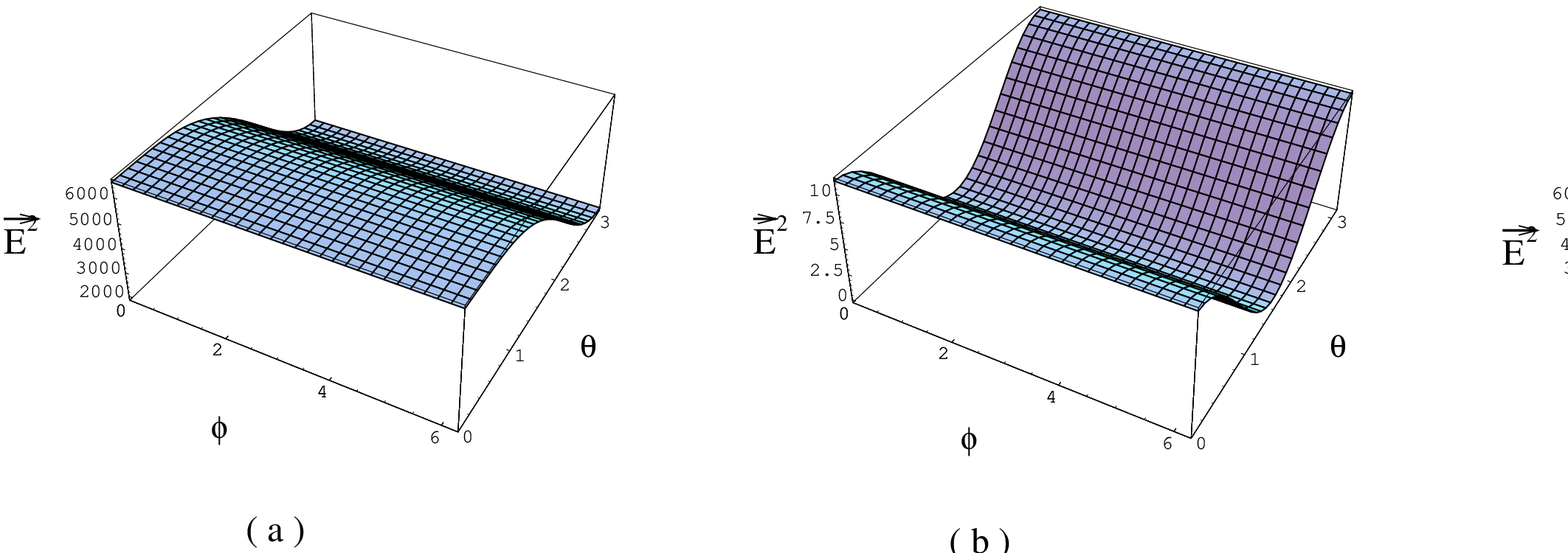}
\caption{} 
\label{} 
\end{figure}

\end{document}